\documentclass[letterpaper,11pt]{article}

\usepackage[utf8]{inputenc} 
\usepackage{lipsum} 

\usepackage[top=2.5cm, bottom=3cm, left=3.5cm, right=3.5cm,
           heightrounded,marginparwidth=1.5cm, marginparsep=1cm]{geometry} 

\usepackage{changepage} 

\usepackage[shortlabels]{enumitem} 

\usepackage[square,numbers,merge,comma,sort&compress]{natbib} 
\makeatletter
\def\NAT@spacechar{\,}  
\makeatother

\usepackage{amsmath,amssymb,amsfonts,amsthm}
\usepackage{mathtools} 

\usepackage{fnpct}
\setfnpct{dont-mess-around} 

\usepackage{slashed,cancel}
\usepackage{comment}   
\usepackage{relsize}   
\usepackage{setspace}  
\usepackage{moresize}  
\usepackage{epsfig}
\usepackage{latexsym}
\usepackage{mathrsfs,calligra,aurical} 
\usepackage{calc}
\usepackage{float}
\usepackage{appendix}
\usepackage{xargs}
\usepackage{extarrows}
\usepackage{empheq}

\usepackage[blocks]{authblk}  

\usepackage[fulladjust]{marginnote}%
\newcommand{\margnote}[2][-0.05]{\marginnote{\setlength\lineskiplimit{-1000pt}\fontsize{7}{7}\sl\selectfont{#2}}[#1\baselineskip]}

\usepackage{graphicx} 

\usepackage[labelsep=colon]{caption}  
\usepackage[labelsep=colon,aboveskip=10pt,belowskip=10pt]{subcaption}
\usepackage{array,multirow,makecell,booktabs}  
\newcolumntype{X}[2]{>{\centering\arraybackslash$}#1{#2\linewidth}<{$}}
\newcolumntype{R}[1]{>{\raggedleft\arraybackslash$}m{#1\linewidth}<{$}}
\newcolumntype{L}[1]{>{\raggedright\arraybackslash}m{#1\linewidth}}

\makeatletter
\renewcommand\mcell@classz{\@classx
   \@tempcnta \count@
   \prepnext@tok
   \@addtopreamble{
      \ifcase\@chnum
         \hfil
         \mcell@agape{\d@llarbegin\insert@column\d@llarend}\hfil \or
         \hskip1sp
         \mcell@agape{\d@llarbegin\insert@column\d@llarend}\hfil \or
         \hfil\hskip1sp
         \mcell@agape{\d@llarbegin \insert@column\d@llarend}\or
         \mcell@agape{$\vcenter
         \@startpbox{\@nextchar}\insert@column\@endpbox$}\or
         \mcell@agape{\vtop
         \@startpbox{\@nextchar}\insert@column\@endpbox}\or
         \mcell@agape{\vbox
         \@startpbox{\@nextchar}\insert@column\@endpbox}%
      \fi
      \global\let\mcell@left\relax\global\let\mcell@right\relax
    }\prepnext@tok}
\makeatother

\makeatletter
\renewcommand\section{\@startsection{section}{1}{\z@}%
                                   {-3.5ex \@plus -1.3ex \@minus -.7ex}%
                                   {2.3ex \@plus.4ex \@minus .4ex}%
                                   {\normalfont\Large\bfseries}}
\renewcommand\subsection{\@startsection{subsection}{2}{\z@}%
                                   {-2.3ex\@plus -1ex \@minus -.5ex}%
                                   {1.2ex \@plus .3ex \@minus .3ex}%
                                   {\normalfont\large\bfseries}}
\renewcommand\subsubsection{\@startsection{subsubsection}{3}{\z@}%
                                   {-2.3ex\@plus -1ex \@minus -.5ex}%
                                   {1ex \@plus .2ex \@minus .2ex}%
                                   {\normalfont\normalsize\bfseries}}
\renewcommand\paragraph{\@startsection{paragraph}{4}{\z@}%
                                   {1.75ex \@plus1ex \@minus.2ex}%
                                   {-1em}%
                                   {\normalfont\normalsize\bfseries}}
\renewcommand\subparagraph{\@startsection{subparagraph}{5}{\parindent}%
                                   {1.75ex \@plus1ex \@minus .2ex}%
                                   {-1em}%
                                   {\normalfont\normalsize\bfseries}}
\makeatother

\usepackage{titletoc}

\addtocontents{toc}{\addvspace{-0.75em}}  

\titlecontents{section}
  [1.25em] {\addvspace{0.7em plus 0pt}\small}
  {\thecontentslabel\hspace{0.75em}}{}
  {\hspace{0.5em}\titlerule*[0.5em]{.}\contentspage}
  [\addvspace{0.0em plus 0pt}]

\titlecontents{subsection}
  [2.75em] {\addvspace{0.075em plus0pt}\fns}
  {\thecontentslabel\hspace{0.75em}}{\thecontentslabel\hspace{0.75em}}
  {\hspace{0.5em}\titlerule*[0.5em]{.}\small\contentspage}
  [\addvspace{0.075em plus 0pt}]

\usepackage{environ}
\makeatletter
\NewEnviron{subalign}[1]{
\begin{subequations}\label{#1}
\begin{align} \BODY \end{align}
\end{subequations}      }
\makeatother
\makeatletter
\newenvironment{subeqs}%
{\begingroup%
\setlength{\abovedisplayskip}{10pt plus 4pt minus 9pt}%
\setlength{\abovedisplayshortskip}{0pt plus 2pt minus 2pt}%
\setlength{\belowdisplayskip}{12pt plus 3pt minus 9pt}%
\setlength{\belowdisplayshortskip}{7pt plus 3pt minus 4pt}%
\begin{subequations}%
}%
{\end{subequations}\ignorespacesafterend%
\endgroup}%
\makeatother

\makeatletter
{\begingroup%
\setlength{\abovedisplayskip}{{#1}pt plus 2pt minus 9pt}%
\setlength{\abovedisplayshortskip}{0pt plus 0pt minus 2pt}%
\setlength{\belowdisplayskip}{{#2}pt plus 3pt minus 9pt}%
\setlength{\belowdisplayshortskip}{7pt plus 3pt minus 4pt}%
\begin{subequations}%
}%
{\end{subequations}\ignorespacesafterend%
\endgroup}%
\makeatother

\makeatletter
{\begingroup%
\setlength{\abovedisplayskip}{{#1}pt plus 3pt minus 9pt}%
\setlength{\abovedisplayshortskip}{0pt plus 3pt}%
\setlength{\belowdisplayskip}{{#2}pt plus 3pt minus 9pt}%
\setlength{\belowdisplayshortskip}{7pt plus 3pt minus 4pt}%
\begin{equation}%
}%
{\end{equation}\ignorespacesafterend%
\endgroup}%
\makeatother

\makeatletter
{%
\begin{equation}%
\begin{split}%
}%
{\end{split}%
\end{equation}\ignorespacesafterend%
}%
\makeatother

\usepackage{xcolor}
\definecolor{Green}{rgb}{0.05, 0.45, 0.25}
\definecolor{dogwoodrose}{rgb}{0.8, 0.1, 0.55}
\definecolor{RRed}{rgb}{0.7, 0.1, 0.525}

\usepackage{bm}  
\usepackage{dsfont}  

\DeclareMathAlphabet{\mathpzc}{OT1}{pzc}{m}{it}
\DeclareMathAlphabet{\mathcal}{OMS}{cmsy}{m}{n}
\DeclareSymbolFontAlphabet{\Scr}{rsfs}
\DeclareMathAlphabet{\mathbold}{U}{BOONDOX-ds}{m}{n}
\SetMathAlphabet{\mathbold}{bold}{U}{BOONDOX-ds}{b}{n}
\DeclareMathAlphabet{\mathcalboondox}{U}{BOONDOX-calo}{m}{n}
\SetMathAlphabet{\mathcalboondox}{bold}{U}{BOONDOX-calo}{b}{n}
\DeclareMathAlphabet{\mathbcalboondox}{U}{BOONDOX-calo}{b}{n}

\newcommand\eqlinkcol{RRed}



\usepackage[breaklinks=true,backref=page]{hyperref}
\hypersetup{
    bookmarks=true,         
    pdfmenubar=true,        
    pdffitwindow=false,     
    pdfpagemode={UseNone},
    pdfstartview={FitH},
    pdfauthor={},     
    pdfsubject={},   
    pdfcreator={},   
    pdfproducer={},  
    colorlinks=true,
    bookmarks=true,
    bookmarksnumbered=true,
    plainpages,
    a4paper,
    linktoc=page,
    citecolor=blue,
    filecolor=black,
    linkcolor=\eqlinkcol,
    urlcolor=Green,
}
\renewcommand*{\backref}[1]{}
\renewcommand*{\backrefalt}[4]{%
\ifcase #1 %
\relax
\or
~{\small [\textsc{p.~\fns{\!#2}}]}
\else
~{\small [\textsc{p.~\fns{\!#2}}]}%
\fi}

\usepackage{footnotebackref}
\usepackage{hypernat} 

\def\+{~+~}
\def\-{~-~}
\def\={~=~}
\newcommand\fns{\footnotesize}

\newcommand\qRq{\quad\Rightarrow\quad}

\newcommand\Real{\textrm{Re}}

\newcommand\eps{\epsilon}
\newcommand\w{\omega}
\newcommand\Id{\mathds{1}}

\newcommand\hgamma{\bar{\gamma}}
\newcommand\e{\text{e}}

\newcommand\Cl{\mathcal{C}_\mathsmaller{\lambda}}
\newcommand\Clp{\mathcal{C}_\mathsmaller{\lambda'}}
\newcommand\FFD{\mathcal{F}_\mathsmaller{\!\textsc{fd}}}
\newcommand\ff{f(\varphi)}
\newcommand\muchem{\bar{\mu}_\mathsmaller{\!\textsc{c}}}
\newcommand\Gphi{\mathcalboondox{G}_{\varphi}}
\newcommand\Gz{\mathcalboondox{G}_z}
\newcommand\WW{\mathcal{W}(\Omega)}
\newcommand\zetaN{\zeta_{{}_\ms{N}}}
\newcommand\IeS{\mathcal{I}_\ms{S}}
\newcommand\vF{v_\textsc{f}}
\newcommand\kdot{k_{\ml{\centerdot}}}
\newcommand\kdotzero{{\kdot}^{\!\!\ms{0}}}

\newcommand{\ms}{\mathsmaller}

\newcommand{\ml}{\mathlarger}

\newcommand{\dd}{\partial}
\newcommandx{\tts}[1]{\text{\textsmaller{#1}}}
\newcommandx{\dm}[1][1=\mu,usedefault]{\partial_{#1}}
\newcommandx{\dmup}[1][1=\mu,usedefault]{\partial^{#1}}
\newcommandx{\subm}[2][1=p,2=A,usedefault]{{#1}_{\!\mathsmaller{#2}}}
\newcommandx{\subt}[2][1=p,2=A,usedefault]{{#1}_\text{\textsmaller{#2}}}
\newcommandx{\supm}[2][1=p,2=A,usedefault]{{#1}^{\!\mathsmaller{#2}}}
\newcommandx{\supt}[2][1=p,2=A,usedefault]{{#1}^\text{\textsmaller{#2}}}
\newcommandx{\subpt}[3][1=p,2=A,3=B,usedefault]{{#1}^\text{\textsmaller{#3}}_\text{\textsmaller{#2}}}
\newcommandx{\subpm}[3][1=p,2=A,3=B,usedefault]{{#1}^{\mathsmaller{#3}}_{\mathsmaller{#2}}}
\newcommandx{\sh}[1][1=\alpha,usedefault]{\sinh\left(#1\right)}
\newcommandx{\ch}[1][1=\alpha,usedefault]{\cosh\left(#1\right)}
\newcommandx{\sech}[1][1=\alpha,usedefault]{\mathrm{sech}\left(#1\right)}
\newcommandx{\cosech}[1][1=\alpha,usedefault]{\mathrm{cosech}\left(#1\right)} \newcommandx{\LCTd}[4][1=\mu,2=\nu,3=\rho,4=\sigma,usedefault]{\eps_{#1#2#3#4}}
\newcommandx{\LCTu}[4][1=\mu,2=\nu,3=\rho,4=\sigma,usedefault]{\eps^{#1#2#3#4}}

\newcommandx{\gmetr}[2][1=\mu,2=\nu,usedefault]{g_{{#1}{#2}}}
\newcommandx{\invgmetr}[2][1=\mu,2=\nu,usedefault]{g^{{#1}{#2}}}
\newcommandx{\spc}[3][1=\mu,2=a,3=b,usedefault]{{\w_{#1}}^{\!\!{#2}{#3}}}
\newcommandx{\Conn}[3][1=\mu,2=\nu,3=\lambda,usedefault]{{\Gamma_{{#1}{#2}}}^{\!\!#3}}
\newcommandx{\viel}[2][1=\mu,2=a,usedefault]{{e_{#1}}^{\!#2}}
\newcommandx{\inviel}[2][1=a,2=\mu,usedefault]{{e_{#1}}^{#2}}
\newcommandx{\vieluu}[2][1=\mu,2=a,usedefault]{e^{#1#2}}
\newcommandx{\Rdduu}[4][1=\mu,2=\nu,3=a,4=b,usedefault]{{R_{{#1}{#2}}}^{{#3}{#4}}}
\newcommandx{\hgamui}[1][1=0,usedefault]{\hgamma^{\mathsmaller{#1}}}
\newcommandx{\hgamdi}[1][1=0,usedefault]{\hgamma_{{}_{#1}}}
\newcommandx{\gamui}[1][1=0,usedefault]{\gamma^{\mathsmaller{#1}}}
\newcommandx{\gamdi}[1][1=0,usedefault]{\gamma_{{}_{#1}}}

\makeatletter
\normalsize
\makeatother

\DeclareFixedFont\trfont{OT1}{phv}{b}{sc}{11}

\title{%
       \vspace{-1.5cm}
       \begin{adjustwidth}{-0.5in}{-0.5in}
       \centering\boldmath\LARGE\bfseries%
       Graphene properties from curved space Dirac equation
       \end{adjustwidth}%
       \bigskip
       }

\author{\small\textsc{Antonio Gallerati}%
\vspace{0.15em}%
}
\affil{%
\makebox[\textwidth][c]{Politecnico di Torino, Dipartimento di Scienza Applicata e Tecnologia, corso Duca degli Abruzzi 24, 10129 Torino, Italy}
}
\affil{Istituto Nazionale di Fisica Nucleare, Sezione di Torino, via Pietro Giuria 1, 10125 Torino, Italy%
}
\affil{\href{mailto:antonio.gallerati@polito.it}{\texttt{antonio.gallerati@polito.it}}%
}

\date{}

\begin{document}

\maketitle


\begin{abstract}
\noindent
A mathematical formulation for particle states and electronic properties of a curved graphene sheet is provided, exploiting a massless Dirac spectrum description for charge carriers living in a curved bidimensional background. In particular, we study how the new description affects the characteristics of the sample, writing an appropriate conductivity Kubo formula for the modified background. Finally, we provide a theoretical analysis for the particular case of a cylindrical graphene sample.
\end{abstract}

\medskip

\tableofcontents



\pagebreak


\section{Introduction}
The Dirac equation is a relativistic wave equation which describes the behaviour of spin 1/2 particles. The reformulation of the Dirac formalism in curved backgrounds is an intriguing field of research due to its considerable applications in high-energy physics, quantum field theory, gravity, cosmology and condensed matter.\par
The above areas of study describe very different aspects of physics, and the considerable gap between different branches is due to the independent formulation and development of the single fields of research. Some instances of exchange of ideas and techniques between different areas are the physics of superfluids \cite{Zurek:1996sj,Volovik:2000ua,bauerle1996laboratory}, the AdS-CFT correspondence \cite{Maldacena:1997re,Witten:1998qj,Gubser:1998bc}, the possible coupling of the gravitational field with the condensate in a superconductor \cite{modanese1996theoretical,agop2000local,Ummarino:2017bvz}, the concept of topological defect \cite{ruutu1996vortex,vilenkin2000cosmic,teo2010topological}, the study of the Unruh effect in ultracold atoms \cite{Rodriguez-Laguna:2016kri,Kosior:2018vgx}.
Some recent developments in material science provide a new connection between condensed matter, high-energy physics and quantum electrodynamics. In particular, the study of the physics of carbon-based materials like graphene opens a window on the possibility of new direct observation of quantum behaviour in the curved background of a solid state system \cite{Vozmediano2010109,cortijo2007effects}.\par
Graphene crystals were first produced in 2004 as two-dimensional, single carbon atom sheets \cite{Novoselov66,novoselov2005twodimato}. Unlike conventional systems, whose charge carriers are described by the Schrödinger equation with effective mass, graphene's charge carriers are characterized by an equation describing two-dimensional massless Dirac-fermions \cite{novoselov2005twodimgas,gusynin2006unusual}. This gives the possibility to study quasi-relativistic particle behaviour at sub-light speed regime \cite{neto2009electronic} and, as a consequence of this, many high-energy particle physics effects can be measured in a solid state system \cite{katsnelson2007graphene,geim2007rise}.
\par
From the perspective of high energy physics, graphene can provide us a real framework to study what is believed to be (as close as possible) a quantum field in a curved spacetime, with measurable effects pertaining to electronic properties of the sample itself. The peculiar structure of a graphene sheet determines a natural description of its properties in terms of massless, relativistic Dirac pseudoparticles \cite{Iorio:2013ifa}. In the honeycomb lattice structure, there are two inequivalent sites per unit cell; the latter distinction is not referred to different kind atoms (since they are all carbons) but is related to their topological inequivalence: that is how the two-component Dirac spinor emerges, the description being resistent to changes of the lattice preserving this aspect of the structure.\par
The relativistic behavior of the charge carriers can be inferred from the dispersion relation between energy and (quasi)momentum. In the momentum space, the valence and the conductivity band touch at inequivalent points, near which the spectrum is linear in the Fermi velocity $\vF$\,: this behavior is expected in a relativistic theory, whereas, in a non-relativistic system, the dispersion relations are usually quadratic%
\footnote{
a graphene sheet is relativistic in the sense of the Fermi velocity as limiting speed
}%
 \cite{Boada:2010sh}.\par
Dirac physics can be realized for our quasiparticles considering low-lying energy excitations ($E<E_a\sim\vF/a$). If we consider energy ranges below $E_a$, the electrons wavelength is large compared to the lattice length $a$, so that these charge carriers see the graphene sheet as a continuum, justifying the quantum description in 2+1 spacetime. Moreover, quasiparticles with large wavelength are sensitive to sheet curvature effects, claiming a quantum field formulation in curved spacetime%
\footnote{
one should also ask the curvature to be small compared to a
limiting maximal curvature related to the lattice length; if not the description cannot be done in terms of a smooth metric, implying also a bending of the very strong $s$ bonds, that is an instance that does not occur}%
 \cite{Iorio:2013ifa}.\par\smallskip
From an experimental point of view, graphene exhibits particular electronic properties which allow the observations of quantum reactions, such as an anomalous quantum Hall effect \cite{zhang2005experimental,kane2005quantum} or strain-induced pseudomagnetic fields \cite{morozov2006strong,levy2010strain}, opening also the possibility to study exotic phenomenons like Unruh effect \cite{Iorio:2011yz}. The particular mechanical properties of graphene sheets have also suggested immediate applications, such as its use in composite materials \cite{stankovich2006graphene}. In addition to this, graphene presents other experimental responses based on the well-known Klein paradox \cite{katsnelson2006chiral,stander2009evidence}: the latter is associated to the particular electronic-transport properties of the graphene sheet, while it was previously discussed only for high-energy phenomena regarding nuclear physics and black holes \cite{gibbons1975vacuum,dombey1999seventy,belgiorno1999quantum}. The description of the electronic spectrum of graphene through the Dirac equation was used to analyse the electron-phonon coupling in large fullerene molecules \cite{gonzalez1992continuum,kolesnikov2006continuum}, the surface states of a topological insulator \cite{lee2009surface}, the appearance and coupling of effective gauge fields in the honeycomb lattice of carbon-based materials \cite{Gonzalez:1992qn,Vozmediano2010109}.
\par\smallskip
A new possible development comes from the study of particular curved configurations of a graphene sheet, the peculiar electronic properties of the material charge carriers derived from the previously discussed massless Dirac description in a curved background \cite{gonzalez1992continuum,Gonzalez:1992qn,gonzalez1993theoretical,osipov2005electronic,kolesnikov2006continuum,morpurgo2006intervalley,lee2009surface}. The choice of the geometry, the corresponding parametrization and the quantization of some physical quantities, can lead to characteristic observable effects. In particular, some optical responses of the graphene sample can be obtained in peculiar ranges of energy, giving rise to new experimental effects and observations.\par\smallskip
In the following, we will apply the formalism of quantum field theory on curved space in order to appropriately describe the Dirac-like spectrum of graphene charge carriers. We will also study the influence of sample structure on the electronic transport, and how optical measurements in graphene sheets are determined by geometry and microscopical disorder. The work is organized as follows.\par
In Sect.~\ref{sec:geom} we will illustrate the construction of the Dirac equation in curved backgrounds, introducing the vielbein formalism and giving an example of parametrization in cylindrical space. Then, in Sect.~\ref{sec:optic} we will define some optical properties of a curved sample, providing a consistent formulation of the Kubo formula for the optical conductivity in a bidimensional background and a correspondent formula for its polarization property.
In Sect.~\ref{sec:layer} we will see how the massless Dirac spectrum can be used to describe particle states and electronic transport in a graphene cylindrical sheet. In particular, we will study how the new description affects the optical properties of the sample, obtaining the new form for the Hamiltonian and the velocity operators in the curved space; they are subsequently used to write the correspondent conductivity Kubo formula for the modified background. We will also provide a theoretical analysis and interpretation of the results, giving some experimental predictions.
Finally, in Sect.~\ref{sec:concl} we will analyse the obtained results and propose some possible future developments.

\section{Dirac equation in curved space} \label{sec:geom}

\subsection{Basics on Dirac equation}
The Dirac equation was the final result of the construction of a relativistic field equation, whose squared wave function modulus could be consistently interpreted as a probability density. To satisfy this requirement, the equation is of first order in time-derivative, while relativistic invariance requires the equation to be first order in space-derivative too \cite{Peskin:1995ev,dauria:2011sr}.
The most general form for this kind of wave equation is%
\footnote{%
here and in the following, we will adopt natural units, where \;$c=\hbar=1$
}:
\begin{equation}  
i\,\frac{\partial\psi}{\partial t}\=
(-i\,\bm{\alpha}^i\partial_i+\bm{\beta}\,m)\;\psi\=\hat{H}\,\psi
\label{direqgen}\;.
\end{equation}
The final explicit form of the above equation must have some properties:
\begin{enumerate}[i.\,,noitemsep,topsep=2.5pt]
\item   it must be Lorentz-covariant, implying Poincaré invariance;
\item  the solution $\psi$ must satisfy the Klein-Gordon equation (mass-shell condition);
\item   a suitable conserved current, written in terms of $\psi$, must have a positive-definite $0$-component, which can be interpreted as the probability density.
\end{enumerate}
To satisfy the above prescriptions, the $\bm{\alpha}$, $\bm{\beta}$ matrices are required to be anticommuting and to square to the identity, namely:
\begin{equation}
\{\bm{\alpha}^i,\,\bm{\alpha}^i\}=2\,\delta^{ij}\,\Id\;,\qquad
\{\bm{\alpha}^i,\,\bm{\beta}\}=0\;;\qquad
(\bm{\alpha}^i)^2=\bm{\beta}^2=\Id\;,
\label{dircond}
\end{equation}
with no summation over $i$ in the third relation.
If one introduces the new set of matrices
\begin{equation}
\gamma^\ms{0} \equiv \bm{\beta}\;,\qquad
\gamma^i \equiv \bm{\beta}\,\bm{\alpha}^i\;,
\end{equation}
the conditions \eqref{dircond} can be rewritten in the form:
\begin{equation}  
\{\gamma^a,\,\gamma^b\}\=2\,\eta^{ab}\,\Id\;,
\label{cliffalg}
\end{equation}
\sloppy
that is usually referred to as Clifford algebra, where ${i=1,\dots,n}$\, while \,${a=0,\dots,n}$\,. The matrix $\eta^{ab}$ is the inverse of the Minkowski flat metric $\eta_{ab}$ that we choose in the mostly minus convention:
\begin{equation}
\eta_{ab} \=
\left(
  \begin{array}{cc}
    1 &  0     \\
    0 & -\Id   \\
  \end{array}
\right) \quad,
\end{equation}
where $\Id\equiv \Id_\ms{n}$ is a $n\times n$ diagonal identity matrix, the value of $n$ depending on the space dimension.
In terms of the above $\gamma^a$ matrices, the previous equation \eqref{direqgen} can be rewritten in the compact form:
\begin{equation}   
(i\,\gamma^a \dm[a] - m\,\Id)\;\psi(x)\=0\;,
\label{direq}
\end{equation}
where, for the sake of notational simplicity, we have omitted the spinorial indices of $\psi\equiv\psi^\alpha$ and $\gamma^a={(\gamma^a)^\alpha}_\beta$\,. Spinors are objects that transform as scalars under the general space-time coordinate transformations and in a spinor representation $\mathcal{R}$ under the local Lorentz group, so that one can write
\begin{equation}   
\psi^{\prime\,\alpha}(x) \=
{\mathcal{R}\big[\Lambda(x)\big]}^{{\!}^\ml{\alpha}}{}_{{\!}_\ml{\beta}}\;\,\psi^\beta(x)\,.
\end{equation}
Using the explicit form of the Lorentz group generators to construct the Pauli-Lubanski operator, it can be easily shown that the particle described by the Dirac equation has spin $s=\frac12$\,.

\subsection{Curved spaces} \label{subsec:curvsp}
Einstein's theory of gravity is based on the symmetry principle of invariance under general coordinate transformations, seen as local space-time transformations generated by the local translation generators. In this background, the gravitational force can be modelled and described from a geometric point of view in terms of the curvature of space-time. Therefore, one must introduce some new tools in order to conveniently describe general relativity and spinorial objects transformation rules (generalized to curved backgrounds). This can be nicely done through the vielbein formalism \cite{D'Auria:1982nx,Castellani:1991et}.

\subsubsection{Vielbein formalism}
Consider a set of coordinates that is locally inertial, so that one can apply the usual Lorentz behaviour of spinors, and imagine to find a way to translate back to the original coordinate frame. More precisely, let $y^a(x_0)$ denote a coordinate frame that is inertial at the space-time point $x_0$: we shall call these the ``Lorentz'' coordinates. Then,
\begin{equation}   
\viel(x) \=
\left. \frac{\partial y^a (x_0)}{\partial x^\mu} \right|_{x=x_0}
\end{equation}
define the so-called \emph{vielbein}. It defines a local set of tangent frames of the space-time manifold and, under general coordinate transformations, it transforms covariantly as
\begin{equation}
{e'_\mu}^{\!a}(x') \=
\frac{\partial x^\nu}{\partial x'^\mu}\,\viel[\nu][a](x)\;,
\end{equation}
while a Lorentz transformation leads to
\begin{equation}
{e'_\mu}^a (x) \= \viel[\mu][b](x)\;{\Lambda^{{\!}^\ml{a}}}_b\;.
\end{equation}
In particular, the space-time metric can be expressed as
\begin{equation}
g_{\mu\nu}(x) \= \viel[\mu][a](x)\;\viel[\nu][b](x)\;\eta_{ab}\;,
\end{equation}
in terms of the Minkowski flat metric $\eta_{ab}$\,.
We can convert the constant $\gamma_a$ matrices of the inertial frame into $\hgamma_\mu$ matrices in the curved frame by the action of the vielbein:
\begin{equation}    \margnote{Curved $\hgamma$ matrices}
\hgamma_\mu(x)\=\viel\,\gamma_a\;,
\end{equation}
while the inverse vielbein $\inviel$ performs the transformation in the other direction. The vielbein thus takes Lorentz (or ``flat'') latin indices to indices in the coordinate basis (or ``curved'') greek indices.\par
The gamma matrices with upper index
\begin{equation}
\hgamma^\mu(x)\=\invgmetr\,\hgamma_\nu\;,
\end{equation}
satisfy the relation:
\begin{equation}
\{\hgamma^\mu,\,\hgamma^\nu\}\=2\,g^{\mu\nu}\,\Id\;,
\label{cliffalgcurv}
\end{equation}
that holds in curved background and is the equivalent form of previous \eqref{cliffalg}.

\paragraph{Covariant derivative, spin connection.}
The choice of the locally inertial frame $y^a$ is unique up to Lorentz transformations given by the Lorentz generators $M_{ab}$\,. In order to couple fields, we can define Lorentz covariant derivatives:
\begin{equation}   
\mathcal{D}_\mu \= \dm + \frac14\,\spc\,M_{ab}\;,
\label{covder}
\end{equation}
where
\begin{equation}
M_{ab} \= \frac12\,[\gamma_a,\gamma_b]\;.
\end{equation}
The $\spc$ object defines the \emph{spin connection}, that can be seen as the gauge field of the local Lorentz group and is determined through the vielbein postulate (tetrad covariantly constant) \cite{green1987superstring}:
\begin{equation}
\mathcal{D}_\mu \viel[\nu] - \Conn\,\viel[\lambda] \= 0 \;,
\end{equation}
written in terms of the affine connection $\Conn$
\begin{equation}
\Conn \= \frac12\,\invgmetr[\sigma][\lambda]
       \left(\dm\gmetr[\nu][\sigma]+\dm[\nu]\gmetr[\mu][\sigma]-\dm[\sigma]\gmetr\right)\;.
\end{equation}
The spin connection can be explicitly written as
\begin{equation}
\spc \= \viel[\nu][a]\,\dm\vieluu[\nu][b]+\viel[\nu][a]\,\Conn[\mu][\lambda][\nu]\,\vieluu[\lambda][b]\;.
\end{equation}

\subsubsection{Example: cylindrical space} \label{subsubsec:cylsp}
For our purposes, it can be instructive to derive an explicit example of parametrization for a geometrical space consisting of a 2D-surface (cylindrically) wrapped around itself several times. The symmetry of the chosen space together with the periodicity on the angular coordinate, will lead to a particular form for the solution of the Dirac equation, in addition to a quantization condition on the transverse component of the electron momentum.\par
The parametrization can be done in terms of three-dimensional spacetime coordinates $x^\mu_{{}_\ms{(3)}}$:
\begin{equation}
x^\textsc{a}_{{}_\ms{(4)}}\equiv x^\textsc{a}
=\big(t,\,x,\,y,\,z\big)
\;\qRq\;
x^\mu_{{}_\ms{(3)}}\equiv x^\mu
=\big(t,\,\varphi,\,z\big) \quad,
\end{equation}
and, using cylindrical coordinates, one can write
\begin{equation}
\begin{split}
x&\=r(\varphi)\;\cos(\varphi)\;,\\
y&\=r(\varphi)\;\sin(\varphi)\;,\\
z&\=z\;,
\end{split}
\end{equation}
where we have considered a possible $\varphi$-dependence for the radius. The Jacobian can be written as
\begin{equation}
J_\mu^{{}^\mathlarger{\textsc{a}}} \=
\frac{\partial x^\textsc{a}}{\partial x^\mu}\=
\left(
  \begin{array}{ccc}
    1  &                        0                                &  0   \\
    0  &  -r(\varphi)\,\sin(\varphi)+r'(\varphi)\,\cos(\varphi)  &  0   \\
    0  &  +r(\varphi)\,\cos(\varphi)+r'(\varphi)\,\sin(\varphi)  &  0   \\
    0  &                        0                                &  1   \\
  \end{array}
\right)        \quad,
\end{equation}
where
$r'(\varphi)\equiv\dfrac{\dd r(\varphi)}{\partial\varphi}$\;.\par\medskip
The metric $\gmetr$ of the curved space has the form:
\begin{equation} 
\gmetr\=\left(J_\mu^{{}^\mathlarger{\textsc{a}}}\right)^{\!\textsc{t}}\eta_\textsc{ab}\;\,J_\nu^{{}^\mathlarger{\textsc{b}}}\=
\left(
  \begin{array}{ccc}
    1   &              0                 &   0    \\
    0   & -r(\varphi)^2-{r'(\varphi)}^2  &   0    \\
    0   &              0                 &  -1    \\
  \end{array}
\right)  \quad,
\end{equation}
so that the line element is written as
\begin{equation}
ds_{{}_\textsc{(c)}}^2\=\gmetr\,dx^\mu dx^\nu
\=dt^2-d\varphi^2\left(r(\varphi)^2+{r'(\varphi)}^2\right)-dz^2\;.
\label{dx2}
\end{equation}
The vielbeins can be easily parameterized as
\begin{equation}
\viel\=\left(
  \begin{array}{ccc}
    1  &  0  &  0    \\
    0  &  -\sqrt{r(\varphi)^2+r'(\varphi)^2}  &  0    \\
    0  &  0  &  -1    \\
  \end{array}
\right)  \quad.
\end{equation}

\paragraph{Gamma matrices.}
In a three-dimensional (flat) spacetime, a possible set of gamma matrices  can be written in terms of the Pauli matrices as
\begin{equation} 
\gamma_a\=\big\{\,\sigma_3\,,\;-i\sigma_1\,,\;i\sigma_2\,\big\}\=
\Biggl\{\;\,
\left(\begin{array}{cc}
           1  &  0  \\
           0  & -1  \\
        \end{array}\right)\,,\;
\left(\begin{array}{cc}
           0  & -i  \\
          -i  &  0  \\
        \end{array}\right)\,,\;
\left(\begin{array}{cc}
           0  &  1  \\
          -1  &  0  \\
        \end{array}\right)\;\,
\Biggr\} \;,\quad
\label{gammaflat}
\end{equation}
and one can easily verify that the matrices $\gamma^a=\eta^{ab}\gamma_b$ satisfy the Clifford algebra \eqref{cliffalg}.\par\smallskip
The curved $\hgamma$ matrices for the cylindrical space can be obtained using the vielbein as:
\begin{equation}
\hgamma_\mu \= \viel\,\gamma_a \=
\left\{\;\,
\left(\begin{array}{cc}
           1  &  0  \\
           0  & -1  \\
        \end{array}\right)\,,\;
\left(\begin{array}{cc}
           0      &  i\;f(\varphi)  \\
    i\;f(\varphi) &       0         \\
        \end{array}\right)\,,\;
\left(\begin{array}{cc}
           0  & -1  \\
           1  &  0  \\
        \end{array}\right)\;\,
\right\} \;, \quad
\label{gammacurv}
\end{equation}
with
\begin{equation}
f(\varphi)\=\sqrt{r(\varphi)^2+{r'(\varphi)}^2}\;.
\end{equation}
The reader can check that the upper indexed gamma matrices ${\hgamma^\mu=\invgmetr\,\hgamma_\nu}$\, satisfy the Clifford algebra \eqref{cliffalgcurv} for curved spaces.

\subsection{Solution of Dirac equation in cylindrical space}
\label{subsec:solcyl}
In the previous section we have seen how to modify the flat $\gamma$ matrices and the derivative term of the Dirac equation to describe the dynamics in curved spaces.\par
Let us now consider the case of the above cylindrical surface of Sect.~\ref{subsubsec:cylsp}. The Dirac equation in curved space is written as:
\begin{equation}  
(i\,\hgamma^\mu\,\mathcal{D}_\mu - m\,\Id)\;\Psi \= 0 \;,
\label{direqcurv}
\end{equation}
where the $\hgamma$ matrices have been explicitly obtained in \eqref{gammacurv} and the covariant derivative has been defined in \eqref{covder}. Because of the particular symmetry of the chosen space, the spin connection vanishes, so that the curved Dirac equation simplifies in:
\begin{equation}
(i\,\hgamma^\mu \dm - m\,\Id)\;\Psi \= 0\;.
\label{direqcyl}
\end{equation}
The solution of the above equation can be written as:
\begin{equation} 
\Psi \= \e^{-i\,\lambda\,E\,t}\,
        \left(\begin{array}{c}
               \Phi_\textsc{a}(\varphi,z) \\
               \Phi_\textsc{b}(\varphi,z) \\  \end{array}\right) \quad,
\label{Psisol}
\end{equation}
where
\begin{equation}
\begin{split}
\Phi_\textsc{a}(\varphi,z)&\=\Cl\;\e^{i\,k_\varphi\,\mathlarger{\int}^{{}^\varphi}{\!\!f(\varphi')\,d\varphi'}}\;\,\e^{i\,k_z\,z}\quad,\\[1ex] \Phi_\textsc{b}(\varphi,z)&\=\Cl\;\e^{i\,k_\varphi\,\mathlarger{\int}^{{}^\varphi}{\!\!f(\varphi')\,d\varphi'}}\;\,\e^{i\,k_z\,z}\;\;\frac{i\,k_\varphi+k_z}{\lambda\,E+m}\quad,
\label{PhiABsol}
\end{split}
\end{equation}
with
\begin{equation}
\Cl=\sqrt{\frac{\lambda\,E+m}{2\,E}}\;,\qquad
f(\varphi)=\sqrt{r(\varphi)^2+{r'(\varphi)}^2}\;,\qquad
\lambda=\pm1\;,\qquad
\label{Cnorm}
\end{equation}
and where the value of $\lambda$ labels hamiltonian eigenstates having positive or negative eigenvalues. The reader can verify that the above solution satisfies the Dirac equation (\ref{direqcyl}) together with the on-shell condition:
\begin{equation}
\quad
E^2\={\vec{k}}^2+m^2\;,
\quad\qquad \vec{k}=(k_\varphi,\,k_z)\;.
\label{onshellcond}
\end{equation}

\paragraph{Quantization of the momentum.}
The periodicity condition on the angular coordinate of the cylindrical surface can be expressed as
\begin{equation}
\varphi=\varphi+2\pi n\;,\qquad
n\in\mathbb{N}\;.
\end{equation}
If we want periodic solutions in $\varphi$, the exponential factor in (\ref{PhiABsol})
\begin{equation}
\e^{i\,k_\varphi\,{\mathlarger{\int}^{{}^\varphi}}{\!\!f(\varphi')\,d\varphi'}}
~\equiv~\e^{i\,\vartheta(\varphi)}\;,
\end{equation}
must satisfy:
\begin{equation}
\vartheta(2\pi)-\vartheta(0)\=2\pi n\;, \qquad\qquad
n\,\in\,\mathds{N}\;.
\end{equation}
This leads to the definition of a geometrical factor $\zeta$
\begin{equation}
\zeta~\equiv~
\int^{{}^{2\pi}}_{{}_0}{\!\!\!\!\!f(\varphi')\,d\varphi'}
\=\frac{2\pi n}{k_\varphi}\;,
\end{equation}
coinciding with the cylinder circumference, from which follows the \emph{quantization condition} for the $k_\varphi$-momentum:
\begin{equation}   
k_\varphi \= \frac{2\pi n}{\zeta}\;.
\label{kquantcond}
\end{equation}


\section{Optical properties}  \label{sec:optic}
Now we can study some physical, measurable properties of a graphene curved surface, whose charge carriers are described by a massless Dirac spectrum, using the mathematical tools developed in the previous Sect.~\ref{subsec:curvsp}. In particular, we will be interested in getting the optical conductivity of a graphene layer and obtaining a general formula for the polarization angle of refractive waves.

\subsection{Conductivity}
Our starting point is the Kubo formula \cite{kubo1956general,kubo1957statistical,mahan2013many,chaves2014optical} adapted to a bidimensional background:
\begin{equation}  \margnote{Kubo formula}
\sigma_{\!\mu\nu}\=
i\;\frac{\IeS}{\Omega}\;\sum_\ms{\alpha,\beta}\,\frac{\FFD(E_\alpha-\muchem)~-~\FFD(E_\beta-\muchem)}{\Omega-\Omega_{\alpha\beta}-i\,\epsilon}\;{v_\mu}^{\!\alpha\beta}\,{v_\nu}^{\!\beta\alpha}\;.
\label{kubo}
\end{equation}
In the above formula, $\FFD(\Upsilon)$ is the Fermi-Dirac function
\begin{equation}
\FFD(\Upsilon)\=\frac{1}{1+\e^{\Upsilon/(k_\textsc{b}T)}}\;,
\end{equation}
\sloppy
$k_\textsc{b}$ is the Boltzmann constant, \,$\muchem$\, is the chemical potential, $\Omega$ is the light energy, ${\Omega_{\alpha\beta}=E_\beta-E_\alpha}$ is the transition energy, $\IeS=\IeS(e,S)$ is a prefactor depending on electric charge of the particle and layer dimension $S$, $\epsilon$ is an infinitesimal quantity. Finally, ${v_\mu}^{\!\alpha\beta}$ are the matrix elements of the velocity operator $\hat{v}_\mu$:
\begin{equation}
{v_\mu}^{\!\alpha\beta}\=\langle \alpha|\,\hat{v}_\mu\,|\beta\rangle\;.
\label{matrixel}
\end{equation}
The operator $\hat{v}_\mu$ can be obtained from the standard quantum mechanics formula
\begin{equation}
\hat{v}_\mu=i\,\left[\hat{H},\,\hat{x}_\mu\right]\;.
\end{equation}
In this regard, the explicit formula of the Hamiltonian is needed. In Sect.\ \ref{sec:layer} we will discuss the case of a graphene cylindrical geometry.

\paragraph{Longitudinal conductivity.}
Let us now focus on the longitudinal optical conductivity, namely:
\begin{equation}
\sigma_{\!\mu\mu}\=
i\,\frac{\IeS}{\Omega}\,\sum_\ms{\alpha,\beta}\frac{\FFD(E_\beta-\muchem)~-~\FFD(E_\alpha-\muchem)}{\Omega-\Omega_{\alpha\beta}-i\,\epsilon}\;{\big|{v_\mu}^{\!\alpha\beta}\big|}^2\;.
\label{sigma}
\end{equation}
For infinitesimal $\epsilon$\,, we have
\begin{equation}
\frac{g(x)}{x+i\,\epsilon}\=\frac{g(x)}{x}-i\,\pi\;\delta(x)\,g(x)
\qquad\; \text{with}\;\epsilon\rightarrow0\;,
\end{equation}
and the real part of the optical conductivity can thus be expressed as:
\begin{equation}   
\textrm{Re}[\sigma_{\mu\mu}]\=
\frac{\pi\,\IeS}{\Omega}\,\sum_\ms{\alpha,\beta}\Delta\!\left(E_\alpha,E_\beta,\muchem\right)\;{\big|{v_\mu}^{\!\alpha\beta}\big|}^2\;\,\delta\!\left(\Omega-\Omega_{\alpha\beta}\right)\,,
\label{Resigma}
\end{equation}
with
\begin{equation}
\Delta\big(E_\alpha,E_\beta,\muchem\big)\=\FFD(E_\beta-\muchem)-\FFD(E_\alpha-\muchem)\;.
\label{Delta}
\end{equation}
The Dirac delta acts as an energy conservation constraint that has to be solved, together with the on-shell condition (\ref{onshellcond}), to fix the momentum values. In fact, one can write in general:
\begin{equation}
\delta\big(\Omega-\Omega_{\alpha\beta}(\kdot)\big)~\equiv~
\delta\big(g(\kdot)\big) \=
\sum_{\kdotzero} \frac{\delta\!\left(\kdot-\kdotzero\right)}{\big|g'(\kdotzero)\big|}\;,
\label{DiracDeltaConstr}
\end{equation}
where we have made explicit the dependence of the energy transition $\Omega_{\alpha\beta}$ on the value of the $\kdot$-component of the momentum, and where the sum is performed over the values $\kdotzero$ that are roots of the Dirac delta argument $g(\kdot)$, i.e.\ the values satisfying the energy conservation constraint $\Omega=\Omega_{\alpha\beta}$.\par\smallskip
One can thus obtain a new form for the real part of the optical conductivity that can be written as:
\begin{equation}
\Real[\sigma_{\mu\mu}]\=
\frac{\pi\,\IeS}{\Omega}\,\sum_\ms{\alpha,\beta}\Delta(E_\alpha,E_\beta,\muchem)\;\,{\big|{v_\mu}^{\!\alpha\beta}\big|}^2\;\,\WW\;\,\delta\!\left(\kdot-\kdotzero\right)\,, \label{ResigmaD}
\end{equation}
with
\begin{equation}
\WW\=\frac{1}{\;\;\left|\dfrac{d\Omega_{\alpha\beta}}{d\kdot}(\kdotzero)\right|}\;\;.
\label{WOmega}
\end{equation}

\subsection{Polarization}
Once the conductivity is given, it is also possible to write, for normal incident light on a graphene layer, the reflection and transmission coefficients as \cite{hanson2008dyadic,hipolito2012enhanced,falkovsky2007optical}
\begin{align}
\mathcal{R}\=&\dfrac{Z_\ms{(2)}-Z_\ms{(1)}-Z_\ms{(1)}\,Z_\ms{(2)}\,\sigma_{\!\mu\mu}}{Z_\ms{(2)}+Z_\ms{(1)}+Z_\ms{(1)}\,Z_\ms{(2)}\,\sigma_{\!\mu\mu}}\;,\\[1.5ex]
\mathcal{T}\=&1+\mathcal{R}\=\dfrac{2\,Z_\ms{(2)}}{Z_\ms{(2)}+Z_\ms{(1)}+Z_\ms{(1)}\,Z_\ms{(2)}\,\sigma_{\!\mu\mu}}\;.
\end{align}
where $Z_\ms{(n)}=\sqrt{\frac{\mu_n}{\epsilon_n}}$ is the impedence of each medium. The reflection and transmission probability (reflectance and transmittance \cite{bohren2008absorption}) are then given by the square of the previous expressions:
\begin{equation}
\bar{\mathpzc{r}}\=\mathcal{R}^2\;, \;\qquad\; \bar{\mathpzc{t}}\=\mathcal{T}^2\;.
\end{equation}


\section{Graphene cylindrical layer}  \label{sec:layer}
Let us consider the previous graphene layer cylindrically wrapped around itself, whose geometrical properties have been described in Sect.~\ref{subsubsec:cylsp}. In what follows, we can safely consider massless particles ($m=0$) \cite{novoselov2005twodimgas,gusynin2006unusual}, to better appreciate the properties due to geometric effects of the curved surface.
\begin{figure}[!htb]
\centering
\includegraphics[width=1\columnwidth]{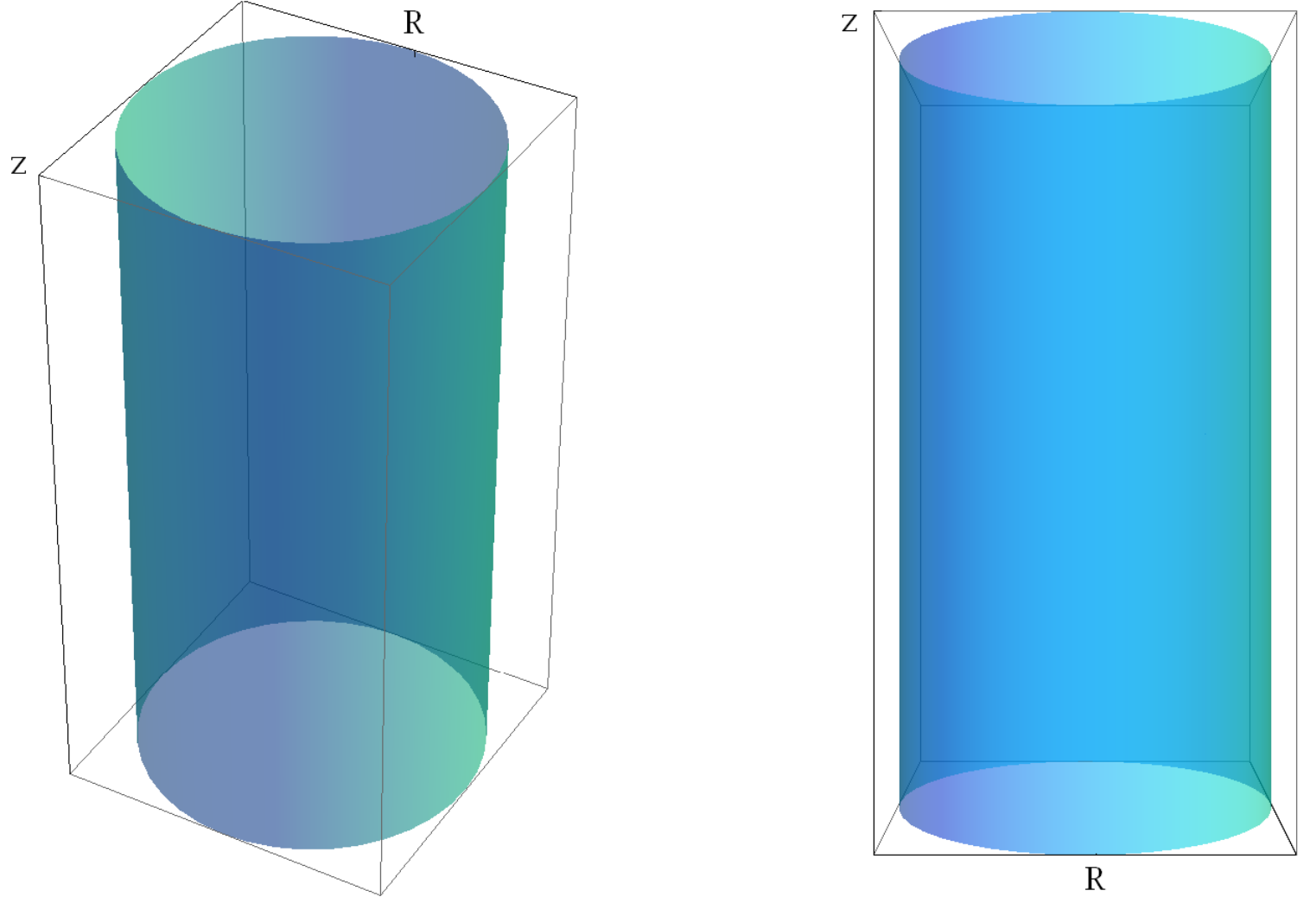}
\label{fig:cyl}
\end{figure}
%

%
%

\subsection{Geometrical analysis}
To find a final expression for the real part of the optical conductivity, we need an explicit expression for the velocity operators matrix element \eqref{matrixel} and for the $\WW$ function of eq.\ \eqref{WOmega}.

\paragraph{Hamiltonian and velocity operators.}
The Hamiltonian of the cylindrical curved graphene can be read from the Dirac equation, taking into account the Schrodinger equation
\begin{equation}
i\,\partial_t\Psi\=\hat{H}\,\Psi\;.
\end{equation}
Using the explicit form of the $\invgmetr$ inverse metric and of the curved gamma matrices, one can isolate the $0$-term of the Dirac equation (\ref{direqcyl}) with $m=0$, obtaining
\begin{equation}
i\,\partial_t\Psi\=i\,\left(\frac{\hgamdi[0]\hgamdi[1]}{\ff^2}\,\dm[\varphi]+\hgamdi[0]\hgamdi[2]\,\dm[z]\right)\,\Psi\;,
\end{equation}
and, therefore, the cylindrical space Hamiltonian operator is obtained as:
\begin{equation} 
\hat{H}_\text{cyl}\=i\,\frac{\hgamdi[0]\hgamdi[1]}{\ff^2}\,\dm[\varphi]+i\,\hgamdi[0]\hgamdi[2]\,\dm[z]\quad.
\end{equation}
Now it is possible to infer the form of the velocity operator from
\begin{equation}
\hat{v}_\mu\=i\,\left[\hat{H}_\text{cyl}\,,\,\hat{x}_\mu\right]\;,
\end{equation}
obtaining:
\begin{subeqs} \label{veloper} 
\begin{align} \margnote{Velocity operators}
\hat{v}_\varphi&\=-\frac{\hgamdi[0]\hgamdi[1]}{\ff^2}\=\frac{\gamdi[0]\gamdi[1]}{\ff}\;,\\[1.5\jot]
\hat{v}_z&\=-\hgamdi[0]\hgamdi[2]\=\gamdi[0]\gamdi[2]\;,
\end{align}
\end{subeqs}
in terms of the flat gamma matrices (\ref{gammaflat}).\medskip

\paragraph{Velocity operator matrix elements.}
The final expression for the matrix elements
\begin{equation}
{v_\mu}^{\!\alpha\beta}\=\langle\,\Psi^{\prime\,\alpha}\,|\,\hat{v}_\mu\,|\,\Psi^\beta\,\rangle
\end{equation}
that appear in the real part of the optical conductivity \eqref{Resigma} is calculated using the explicit solution $\Psi^\alpha$ given in \eqref{Psisol}-\eqref{Cnorm} and the velocity operators obtained in \eqref{veloper}.\par
We decide to label the solution $\Psi^\alpha$ in terms of the natural number $n$, related to the quantization condition \eqref{kquantcond}, and Hamiltonian label eigenvalue $\lambda$:
\begin{equation}
\Psi^\alpha~\equiv~\Psi^{(\lambda,n)}\quad,
\label{Psilab}
\end{equation}
and obtain, for the $\hat{v}_\varphi$ operator, the matrix element
\begin{equation}
{v_\varphi}^{\!\alpha\beta}\=
2\pi\;\Cl\,\Clp\left(\frac{-i\,k_z+k_\varphi}{\lambda\,E}+\frac{i\,k_z+k_\varphi'}{\lambda'\,E'}\right)\;
\Gphi(n,n')\;\;,
\label{vphiME}
\end{equation}
with
\begin{equation}
\Gphi(n,n')\=\int^{{}^{2\pi N}}_{{}_0} {\!\!\!d\varphi\;\;\frac{\e^{i\,\tfrac{2\pi(n-n')}{\zetaN}
\mathlarger{\int}^{{}^{\varphi}}\!\!\!d\varphi'f(\varphi')}}{\ff}}\quad,
\label{Gphi}
\end{equation}
and
\begin{equation}
\zetaN~\equiv~
\int^{{}^{2\pi N}}_{{}_0}{\!\!\!\!\!f(\varphi')\,d\varphi'}\;.
\end{equation}
The number $N$ takes into account the wingdings of the graphene layer wrapped around itself \cite{xie2009controlled}. From a physical point of view, however, it acts as a momentum cut-off, just limited by the validity of the (long wavelength) continuum approximation. This means that the number $N$ must be small when compared with the circumference of the cylinder divided graphene lattice spacing ($N\ll 2\pi R/a$). \par\smallskip
\sloppy
The matrix element is used within the formula (\ref{ResigmaD}) to obtain the real part of the optical conductivity, together with the constraint ensured by the presence of the Dirac delta. The solution of the latter, i.e.\ $\Omega=\Omega_{\alpha\beta}(k_z)$, fixes the value of the $k_z$ component in the massless case to the value $k_z^\ms{0}$ (see formula \eqref{DiracDeltaConstr}):
\begingroup
\belowdisplayshortskip=10pt
\begin{equation}
k_z \;\;\rightarrow\;\; k_z^\ms{0}\=
\sqrt{\frac{\Big(k_\varphi^2-k_\varphi^{\prime\,2}\Big)^2-2\,\Big(k_\varphi^2+k_\varphi^{\prime\,2}\Big)\,\Omega^2+\Omega^4}{4\,\Omega^2}}\;\;.
\label{kz0}
\end{equation}
\endgroup
where, for the $k_\varphi$-component, we use the adapted version of (\ref{kquantcond}):
\begin{equation}
k_\varphi\=\frac{2\pi n}{\zetaN}\;,\qquad\;
k_\varphi'\=\frac{2\pi n'}{\zetaN}\;.
\label{kphi}
\end{equation}
In the same way, the matrix element for the $\hat{v}_z$ operator is calculated and reads:
\begin{equation}
{v_z}^{\!\alpha\beta}\=
2\pi\;\Cl\,\Clp\left(\frac{i\,k_\varphi+k_z}{\lambda\,E}+\frac{-i\,k_\varphi'+k_z}{\lambda'\,E'}\right)\;
\Gz(n,n')\;\;,
\label{vzME}
\end{equation}
with
\begin{equation}
\Gz(n,n')\=\int^{{}^{2\pi N}}_{{}_0} {\!\!\!d\varphi\;\;\e^{i\,\tfrac{2\pi(n-n')}{\zetaN}
\mathlarger{\int}^{{}^{\varphi}}\!\!\!d\varphi'f(\varphi')}}\;\;\;,
\label{Gz}
\end{equation}
where, again, $k_z$ is fixed to the value $k_z^\ms{0}$ and $k_\varphi$ and $k_\varphi'$ are those of eq.\ \eqref{kphi}.\par
The explicit form of the layer-surface parametrization given in Sect.\ \ref{subsubsec:cylsp} can be used to calculate, through the factor $f(\varphi)$, the functions \,$\Gphi(n,n')$,\, $\Gz(n,n')$\, that modulate the amplitude of the velocity operator matrix elements, and obtain an explicit result for the real part of the optical conductivity (\ref{ResigmaD}).\par
The $\WW$ function of eq.\ \eqref{WOmega}, in the massless case, has the explicit form:
\begin{equation}
\WW\= \left(k_z^\ms{0}\right)^{-1} \left(\frac{1}{\lambda\,E_\ms{0}}+\frac{1}{\lambda'\,E'_\ms{0}}\right)^{-1}
\;;\qquad\quad
E_\ms{0} ~\equiv~ E\big|_\ms{k_z = k_z^\ms{0}}\;\;.\quad
\label{WOm}
\end{equation}
Finally, we obtain a new compact expression for the real part of the longitudinal optical conductivity \eqref{Resigma} in cylindrical geometry, with explicit form:
\begin{subeqs} \label{Resigmaphiphizz} 
\begin{align}
\Real[\sigma_{\!\ms{\varphi\varphi}}]\=
\frac{\pi\,\IeS}{\Omega}\;\sum_\ms{\alpha,\beta}\,\Delta(E_\alpha,E_\beta,\muchem)\;\;\WW\;\;{\big|{v_\varphi}^{\!\alpha\beta}\big|}^2\;,\\[1ex]
\Real[\sigma_{zz}]\=
\frac{\pi\,\IeS}{\Omega}\;\sum_\ms{\alpha,\beta}\,\Delta(E_\alpha,E_\beta,\muchem)\;\;\WW\;\;{\big|{v_z}^{\!\alpha\beta}\big|}^2\;,
\end{align}
\end{subeqs}
\sloppy
where the function $\Delta(E_\alpha,E_\beta,\muchem)$ is defined in eq.\ \eqref{Delta}. As in formula \eqref{Psilab}, the labels \,${\alpha=(\lambda',n')}$\, and \,${\beta=(\lambda,n)}$\, are related to Hamiltonian energy eigenvalues ${\lambda,\lambda'=\pm1}$ and natural numbers \,$n,n'\in [1,N]$\,.

\subsection{Analysis and experimental predictions}\label{subsec:analysis}
The structure of formula \eqref{Resigmaphiphizz} determines the behaviour of the optical conductivity for our graphene cylindrical layer.\par\smallskip
The functions $\Gphi$, $\Gz$ are geometrical factors related to the form of the manifold describing the graphene sample, and modulate the amplitude of $\Real[\sigma]$. In the case under consideration, they only depend on the number of wingdings of the graphene layer and on the chosen parametrization $r(\varphi)$ of the surface.\par\smallskip
The function $\WW$ of eq.\ \eqref{WOm} exhibits a pole at the energy values $\Omega_0$ for which $k_z^\ms{0}$ vanishes:
\begingroup
\belowdisplayshortskip=10pt
\begin{equation}
k_z^\ms{0}=0  \;\qRq\;
\Omega_0=\left|\frac{2\pi(n\pm n')}{\zetaN}\right|\;,\qquad
n+n'\leq N\;,
\label{poles}
\end{equation}
suggesting a certain number of singularities for the optical conductivity, leading to a possible peak structure%
\footnote{
singularities arising from $\WW$ in the real graphene sample are partially removed by internal disorder such as carrier density inhomogeneity and imperfections of the sample itself, leaving only a multiple peak configuration for the conductivity}%
\footnote{
internal disorder of the sample can be simulated, in the calculations, by performing an ensemble averaging over a distribution of the model parameters \cite{hipolito2012enhanced}}%
.
\par\smallskip
In the case of \emph{intraband transition}, that is between states with same energy eigenvalues $(\lambda \lambda'=1)$, the factor $\Delta\left[E_\alpha,E_\beta,\muchem\right]$ strongly suppresses in general the value of $\Real[\sigma^\tts{intra}]$, according to Pauli exclusion principle. In particular, in case of flat graphene sheet ($n=n'$), the intraband conductivity would vanish. However, in a curved graphene sample, we can find intraband transition with $n\neq n'$, that is between different quantum states. This means that $\Real[\sigma^\tts{intra}]$ is a suppressed quantity, except for transition having $n\neq n'$ and energy values $\Omega\approx\Omega_0$ approaching the poles of the $\WW$ function. We therefore expect a narrow peak structure for the real part of the optical conductivity for small values of $\Omega$, the peaks coinciding with the poles of eq.\ \eqref{poles}.\par\medskip
In the case of \emph{interband transition}, i.e.\ transition between states with different energy eigenvalues $(\lambda \lambda'=-1)$, the matrix element ${\big|v_\varphi^{\tts{inter}}\big|}$ vanishes for $\Omega=\Omega_0$, giving us:
\begin{equation}
\lim_{\Omega \to \Omega_0} \,{\big|v_\varphi^{\tts{inter}}(\Omega)\big|}^2\;\WW\=0\;,
\end{equation}
cancelling the poles of the $\WW$ function and leading to a minimum for the optical conductivity $\Real[\sigma_{\!\ms{\varphi\varphi}}^\tts{inter}]$. The latter is therefore expected to reflect the quantization process \eqref{kphi} of the $k_\varphi$-momentum with a number of oscillation related to the number $N$, i.e.\ to the geometry of our sample.
On the contrary, the matrix element ${\big|v_z^{\tts{inter}}\big|}$ is different from zero for $\Omega=\Omega_0$, leaving the poles of the $\WW$ function: this suggest again a strongly peaked configuration for $\Real[\sigma_{\!\ms{zz}}^\tts{inter}]$.
\par\smallskip
Obviously, in both interband and intraband case, the conductivity will be suppressed for very large values of $\Omega$.\par

\paragraph{Example.}
Let us consider the following parametrization for the cylindrical geometry of our graphene sample:
\begin{equation}
r(\varphi)\=R\,{\left(1-\frac{1}{2\pi}\,\frac{\varphi}{D N}\right)}^{\!2}\;,
\end{equation}
where the coefficient $D$ determines the distance between different layers and $R$ is a typical dimension for the average radius of the cylinder.\par
\begin{figure}[H]
\centering
\begin{adjustwidth}{-2cm}{-1.5cm}
\begin{subfigure}{.8\columnwidth}
  \centering
  \includegraphics[width=.8\columnwidth]{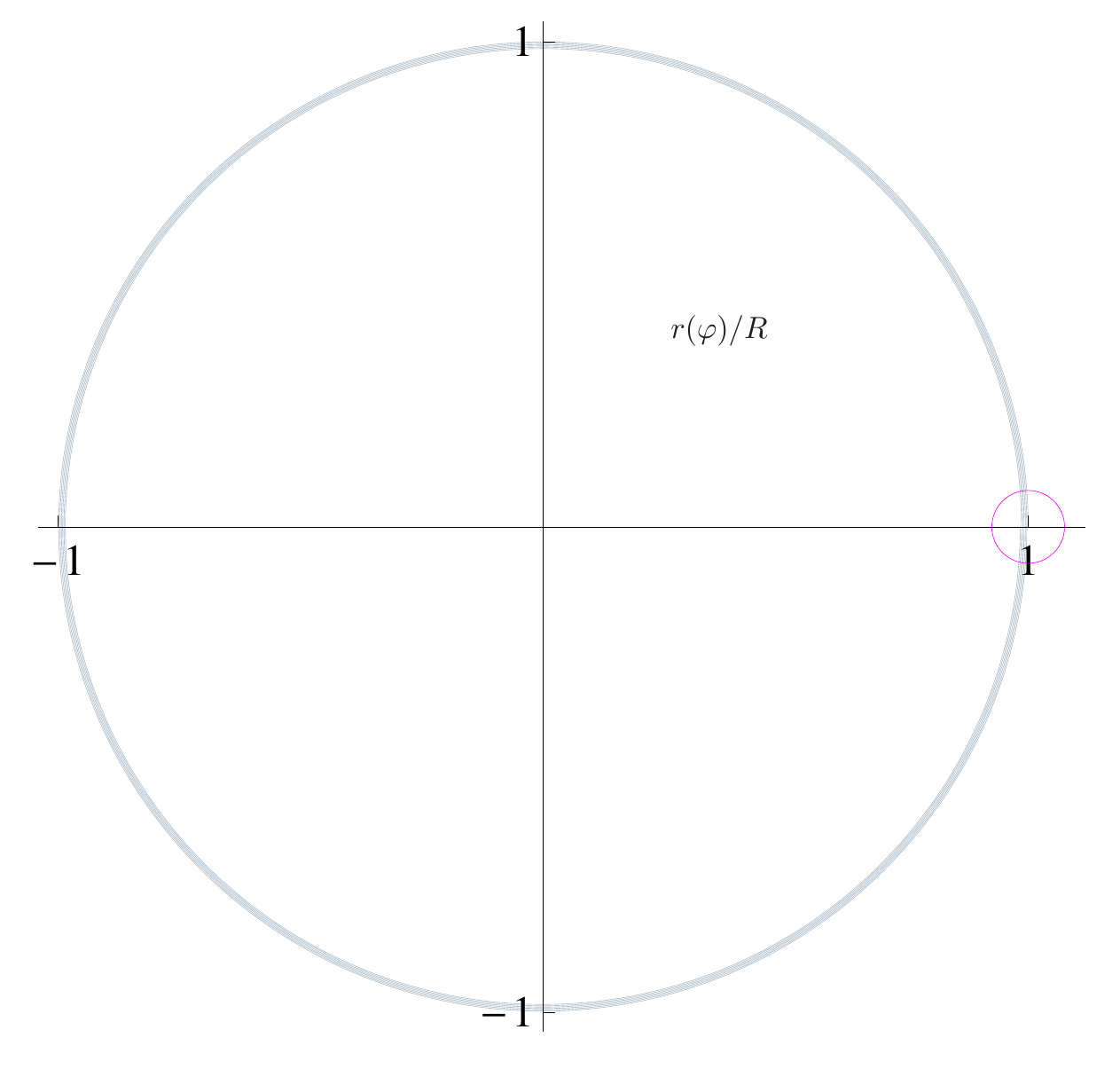}
  \label{subfig:rphi}
\end{subfigure}%
\begin{subfigure}{.45\columnwidth}
  \centering
  \includegraphics[width=.45\columnwidth]{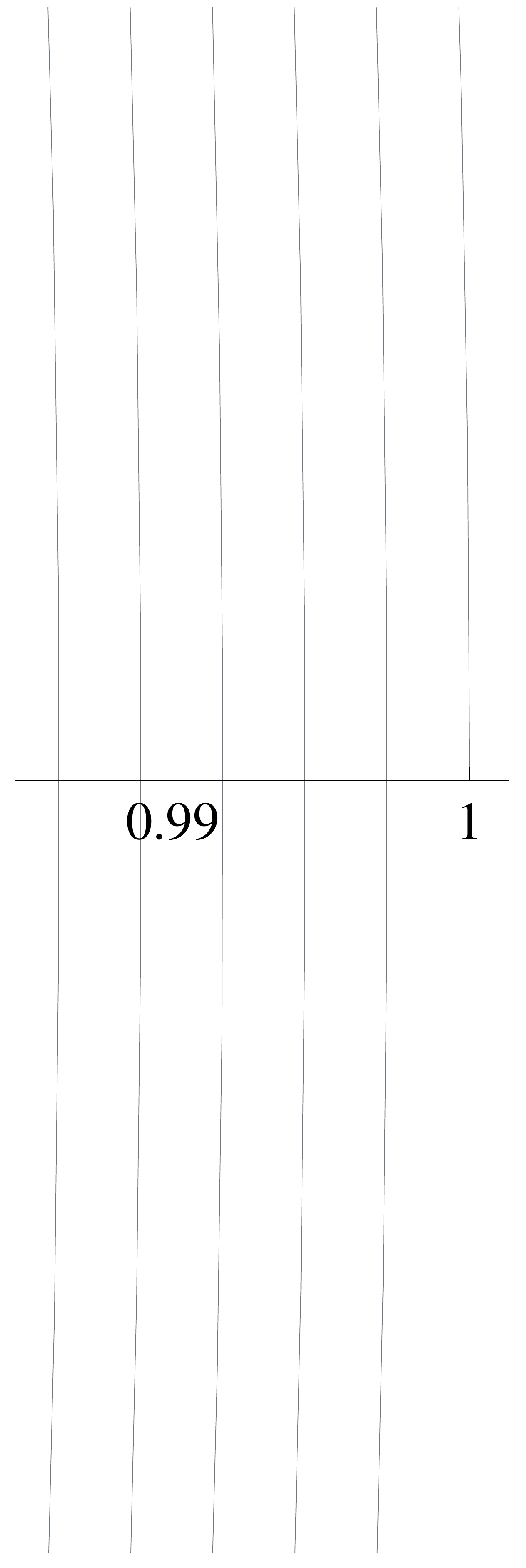}
  \label{fig:rphizoom}
\end{subfigure}
\caption{Parametric plot for \,$\dfrac{r(\varphi)}{R}$\, with \,$D=120$}
\end{adjustwidth}
\label{fig:cylpar}
\end{figure}
The functions $\Gphi$ and $\Gz$ can be explicitly obtained using \eqref{Gphi} and \eqref{Gz} once fixed the number of wingdings and the parameter $D$.
\begin{figure}[H]
\centering
\includegraphics[width=0.6\columnwidth]{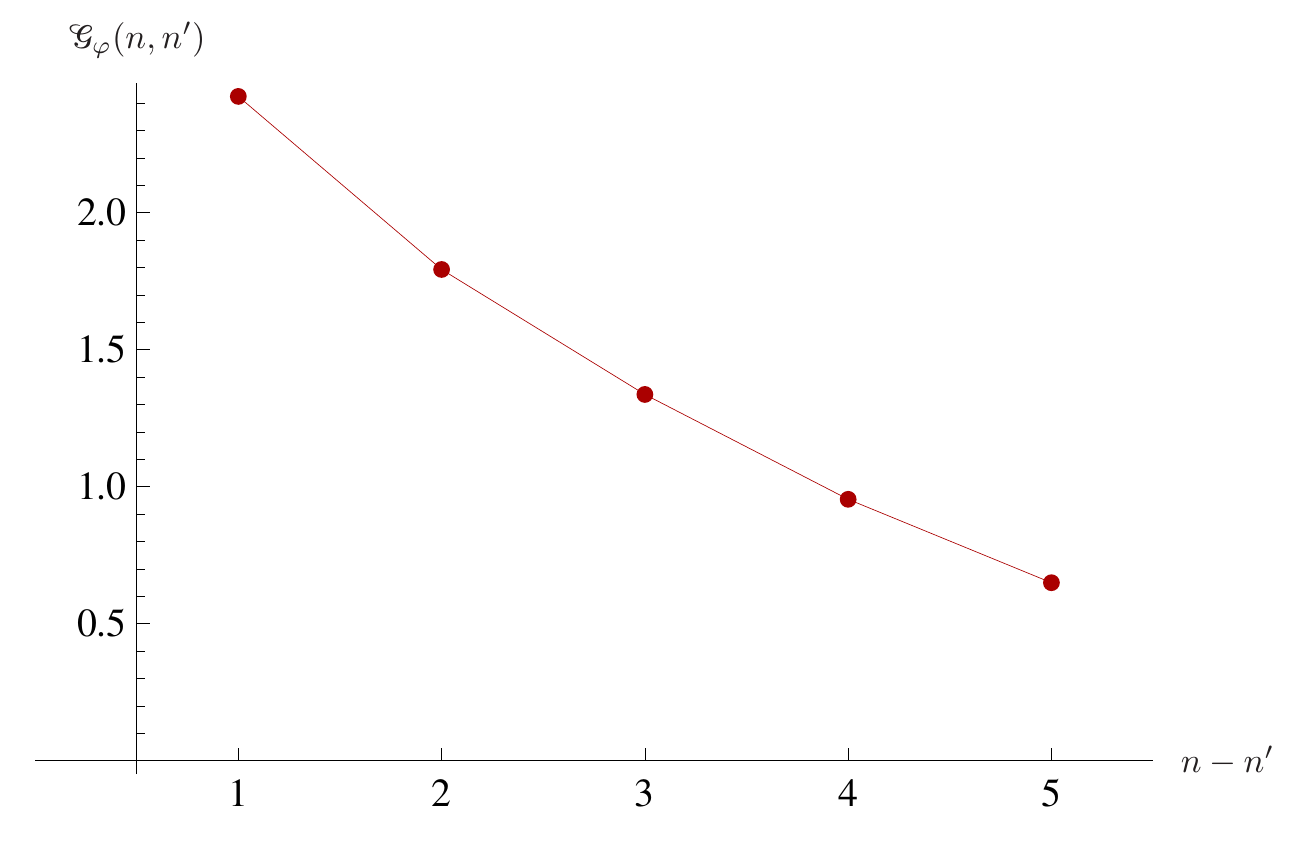}
\caption{Example of the \,$\Gphi(n,n')$\, function for $N=6,\,D=120$\,.}
\label{fig:GGphi}
\end{figure}
\begin{figure}[H]
\centering
\includegraphics[width=0.6\columnwidth]{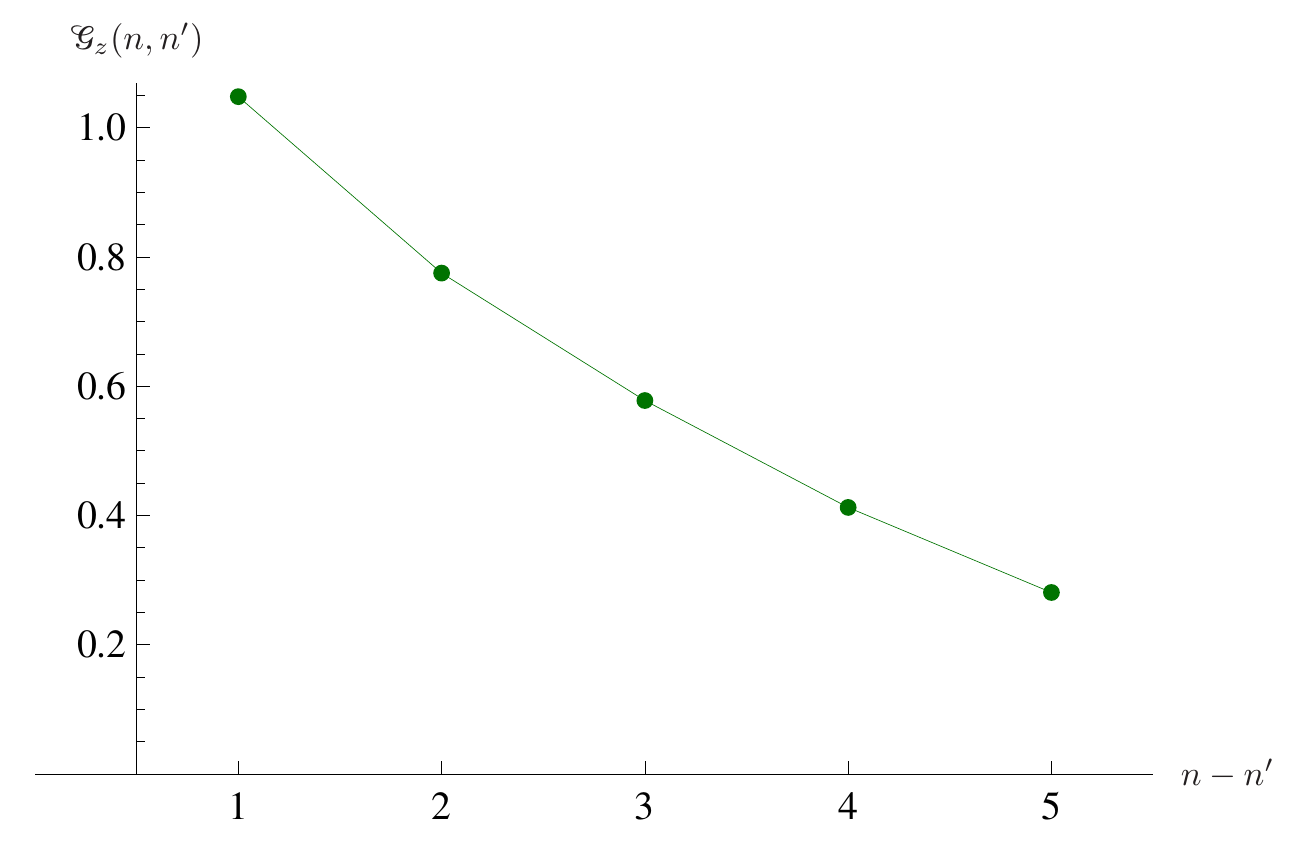}
\caption{Example of the \,$\Gz(n,n')$\, function for $N=6,\,D=120$\,.}
\label{fig:GGz}
\end{figure}
The poles of the $\WW$ function can be found using \eqref{poles} and, choosing $R=90\,\mathrm{nm}$, $N=6$ and $D=120$, one obtains:
\begin{equation}
\Omega_0\=p\,\times\,0.368\,\mathrm{eV}\;, \;\qquad p=1,2,3,\dots,12\;.
\end{equation}
This means that we can expect significant optical responses from our graphene sample in a range of energy $[0.3\div4.5]$ eV, that is, an interval including the visible light energy range.

\section{Conclusions}\label{sec:concl}
We have seen how the curved space formalism can be used to describe the peculiar massless Dirac spectrum of graphene charge carriers. In particular, we focused on a particular sample geometry, having cylindrical symmetry. The resulting formulation leads to a modified Dirac equation, whose solution is used to obtain the matrix transition elements for the velocity operators. The particular geometry of the manifold, represented by the graphene sheet, gives also a suitable quantization rule for the momentum component $k_\varphi$. Finally, the matrix elements define a final expression for the Kubo conductivity, and its simple form can be used for experimental predictions to be tested.\par
In Section \ref{sec:layer} we have seen how the obtained Kubo formula can be used to infer the optical properties of the cylindrical graphene sample and, in Section \ref{subsec:analysis}, we have also shown how a suitable choice for the parametrization can result in a physical response of the sample in particular ranges of energy, including the visible light interval. In this regard, experimental observations are currently taking place.\par
The whole procedure applies in general to different graphene geometries, and can be  suitably and simply modified to describe various symmetric spaces describing the curved graphene sheet.\par

\vspace{1cm}

\section*{\large Acknoledgments}
\vspace{-5pt}
I would like to thank professors Francesco Laviano, Mario Trigiante and Giovanni Ummarino for their extremely helpful advices and support during the preparation of this paper.
The final publication is available at Springer via \href{https://doi.org/10.1140/epjp/i2019-12610-6}{https://doi.org/10.1140/epjp/i2019-12610-6}


\appendix
\addtocontents{toc}{\protect\addvspace{2.5pt}}%
\numberwithin{equation}{section}%
\numberwithin{figure}{section}%


\newpage

\hypersetup{linkcolor=blue}
\phantomsection 
\addcontentsline{toc}{section}{References} 
\bibliographystyle{mybibstyle}
\bibliography{bibliografia} 

\end{document}